\documentclass[12pt]{article}
\usepackage{amssymb}
\usepackage{epsf}
\usepackage{epsfig}
\usepackage{cite}
\newcommand{\lsim}{
\mathrel{\hbox{\rlap{\hbox{\lower4pt\hbox{$\sim$}}}\hbox{$<$}}}}

\newcommand{\gsim}{
\mathrel{\hbox{\rlap{\hbox{\lower4pt\hbox{$\sim$}}}\hbox{$>$}}}}

\begin{document}


\thispagestyle{empty}

\begin{flushright}
CERN-PH-TH/2008-016\\
\end{flushright}

\vspace{2.0truecm}
\begin{center}
\boldmath
\large\bf Prospects for $B$-Decay Studies at the LHC
\unboldmath
\end{center}

\vspace{0.9truecm}
\begin{center}
Robert Fleischer\\[0.1cm]
{\sl Theory Division, Department of Physics, CERN\\
CH-1211 Geneva 23, Switzerland}
\end{center}

\vspace{0.9truecm}

\begin{center}
{\bf Abstract}
\end{center}

{\small
\vspace{0.2cm}\noindent
In this decade, there are huge efforts to explore $B$-meson decays, which 
offer interesting probes to test the quark-flavour structure of the Standard
Model and to search for signals of new physics. Exciting new perspectives
for these studies will soon arise at the LHC, where decays of $B^0_s$ mesons 
will be a key target of the $B$-physics programme. We will discuss theoretical 
aspects of various benchmark channels and address the question of how much 
space for new-physics effects in their observables is left by the recent experimental results from the $B$ factories and the Tevatron. 
}

\vspace{0.9truecm}

\begin{center}
{\sl Invited talk at 3rd High-Energy Physics Conference in Madagascar
(HEP-MAD 07), Antananarivo, Madagascar, 10--15 September 2007\\
To appear in the Proceedings}  
\end{center}

\vfill
\noindent
January 2008

\newpage
\thispagestyle{empty}
\vbox{}
\newpage
 
\setcounter{page}{1}


\section{Introduction}
In the Standard Model (SM), the phenomenon of CP violation can be accommodated 
in an efficient way through a complex phase entering the quark-mixing matrix,
which governs the strength of the charged-current interactions of the quarks 
\cite{KM}. This Kobayashi--Maskawa (KM) mechanism of CP violation is the 
subject of detailed investigations in this decade. The main interest in the study
of CP violation and flavour physics in general is due to the fact that new physics 
(NP) typically leads to new patterns in the flavour sector. This is actually the case in 
several specific extensions of the SM, such as SUSY scenarios, left--right-symmetric 
models, models with extra $Z'$ bosons, scenarios with extra dimensions, or 
``little Higgs" models. Moreover, also the observed neutrino masses point towards an
origin lying beyond the SM \cite{kus}, raising the question of having CP violation 
in the neutrino sector and its connection with the quark-flavour physics. Finally, the
baryon asymmetry of the Universe also suggests new sources of CP violation. These
could be associated with very high energy scales, where a particularly interesting
scenario is provided by ``leptogenesis" \cite{BPY}, involving typically new 
CP-violating sources in the decays of heavy Majorana neutrinos. On the other hand, 
new CP-violating effects arising in the NP scenarios listed above could in fact be 
accessible in the laboratory. 

Before searching for signals of NP, we have first to understand the SM picture.
Here the key problem is due to the impact of strong interactions, leading to
``hadronic" uncertainties. The $B$-meson system is a particularly promising 
probe for the testing of the quark-flavour sector of the SM, and will be the focus 
of this presentation. Decays of $B$ mesons are studied at two kinds of 
experimental facilities. The first are the ``$B$ factories" at SLAC and KEK with
the BaBar \cite{hawkes} and Belle \cite{trabelsi} experiments, respectively. These 
machines are asymmetric 
$e^+e^-$ colliders that have by now produced altogether ${\cal O}(10^9)$ $B\bar B$ pairs, establishing CP violation in the $B$ system and leading to many other 
interesting results. There are currently discussions of a super-$B$ factory, with an 
increase of luminosity by two orders of magnitude \cite{superB}. 
Since the $B$ factories are
operated at the $\Upsilon(4S)$ resonance, only $B^0_d\bar B^0_d$ and 
$B^+_uB^-_u$ pairs are produced. On the other hand, hadron colliders produce, 
in addition to $B_d$ and $B_u$,  also $B_s$ 
mesons,\footnote{Recently, data were taken by Belle at $\Upsilon(5S)$, allowing also access to $B_s$ decays \cite{Belle-U5S}.} as well as $B_c$ and $\Lambda_b$ 
hadrons, and the Tevatron experiments CDF and D0 have reported first 
$B_{(s)}$-decay results. The physics potential of the $B_s$ system can 
be fully exploited at the LHC, starting operation in the summer of 2008. 
Here ATLAS and CMS can also address some $B$-physics topics, although 
these studies are the main target of the dedicated LHCb experiment \cite{pel}.
The central target of these explorations is the well-known unitarity triangle (UT) of the
Cabibbo--Kobyashi--Maskawa (CKM) matrix with its three angles $\alpha$, $\beta$
and $\gamma$, and strongly suppressed ``rare" decays of $B$ mesons. 

The key processes for the exploration of CP violation are non-leptonic decays 
of $B$ mesons, where only quarks are present in the final states. In these transitions,
CP-violating asymmetries can be generated through interference effects. Depending
on the flavour content of their final states, non-leptonic $B$ decays receive 
contributions from tree and penguin topologies, where we distinguish between
QCD and electroweak (EW) penguins in the latter case. The calculation of the
decay amplitudes, which can be written by means of the operator product 
expansion as follows \cite{BBL}:
\begin{equation}
A(B\to f)\sim \sum\limits_k
\underbrace{C_{k}(\mu)}_{\mbox{pert.\ QCD}} 
\times\,\,\, \underbrace{\langle f|Q_{k}(\mu)|B\rangle}_{\mbox{``unknown''}},
\end{equation}
remains a theoretical challenge, despite interesting recent progress through
QCD factorization \cite{QCDF}, PQCD \cite{PQCD}, SCET \cite{SCET}, and 
QCD sum rule applications \cite{QCDSR}.

For the exploration of CP violation, the calculation of the hadronic matrix elements 
$\langle f|Q_{k}(\mu)|B\rangle$ of local four-quark operators can actually be 
circumvented. This feature is crucial for a stringent testing of the CP-violating
flavour sector of the SM. From a practical point of view, two main avenues 
are offered:
\begin{itemize}
\item Amplitude relations allow us in fortunate cases to eliminate the hadronic 
matrix elements. Here we distinguish between exact relations, 
using pure ``tree'' decays  of the kind $B^\pm\to K^\pm D$ \cite{gw,ADS} or 
$B_c^\pm\to D^\pm_s D$ \cite{fw}, and relations, which follow from the flavour symmetries of strong interactions, i.e.\ isospin or $SU(3)_{\rm F}$, and 
typically involve $B_{(s)}\to\pi\pi,\pi K,KK$ modes~\cite{GHLR}. 
\item In decays of neutral $B_q$ mesons ($q\in\{d,s\}$), the interference between
$B^0_q$--$\bar B^0_q$ mixing and $B^0_q, \bar B^0_q\to f$ decay processes 
leads to ``mixing-induced" CP violation. If one CKM amplitude dominates the
decay, the essentially ``unknown" hadronic matrix elements cancel. The key 
application of this important feature is the measurement of $\sin2\beta$ through the 
``golden" decay $B^0_d\to J/\psi K_{\rm S}$ \cite{bisa}.  
\end{itemize}

Following these lines, various processes and strategies emerge for the exploration
of CP violation in the $B$-meson system (for a more detailed discussion, see
\cite{RF-lect}). In particular, decays with a very different dynamics allow us to
probe the same quantities of the UT. These studies are complemented by rare 
decays of $B$ and $K$ mesons, which originate from loop processes in 
the SM model and show interesting correlations with the CP violation in the $B$ system. 
In the presence of NP, discrepancies should show up in the resulting roadmap of 
quark-flavour physics at some level of accuracy.

\boldmath
\section{A Brief Look at the $B$-Factory Data}\label{sec:2}
\unboldmath
Comprehensive and continuously updated analyses of the UT are performed by
the ``CKM Fitter Group'' \cite{CKMfitter} and the ``UTfit collaboration'' \cite{UTfit}.
The current data show impressive global agreement with the KM mechanism. 
Nevertheless, there are also potential deviations from the SM description
of CP violation, and LHCb will soon allow us to enter a territory of the 
$B$-physics landscape that is still largely unexplored. 

If a given decay is dominated by SM tree processes, we have typically small effects
through NP contributions to its transition amplitude. On the other hand, we may 
have potentially large NP effects in the penguin sector through new particles in 
the loops or new contributions at the tree level (this may happen, for instance, 
in SUSY or models with extra $Z'$ bosons). The search for such signals of NP 
in the $B$-factory data has been a hot topic for several years, which is reflected
by the great attention that the ``$B\to \pi K$ puzzle" has received 
(see, e.g., \cite{BpiK-papers}). For the CP-averaged branching ratios, the
$B$-factory data have moved towards the SM prediction, while the mixing-induced
CP violation in $B^0\to \pi^0 K_{\rm S}$ may still indicate a deviation from the
SM, which could be accommodated through a modified EW penguin sector
with a large CP-violating phase \cite{BFRS,FRS-07}. This effect is complemented 
by the $B$-factory measurements of the mixing-induced CP asymmetries of other 
penguin-dominated $b\to s$ modes \cite{HFAG}, which can be converted into 
$\sin 2\beta$; an outstanding example is the decay $B^0\to \phi K_{\rm S}$  
\cite{RF-EWP-rev,growo,FM-BphiK}. The corresponding patterns in the data could be 
footprints of the same kind of NP. 

Unfortunately, it is unlikely that the current $B$ factories will allow us to 
establish -- or rule out -- the tantalizing option of having NP in the $b\to s$ 
penguin processes. However, at LHCb, this exciting topic
can be explored with the help of the decay $B^0_s\to \phi\phi$ \cite{FG-1}. 
A handful of events have been observed in this mode a few 
years ago by the CDF collaboration at the Fermilab Tevatron, corresponding to a 
branching ratio of  $(14^{+6}_{-5}\pm 6)\times 10^{-6}$~\cite{Acosta:2005eu}. 
A  proposal for studying time and angular dependence in this decay mode has been
made by the LHCb collaboration \cite{LHCb-Bsphiphi}. The proposal is based on 
an estimated sample of about 3100 events collected in one year of running. In order
to control hadronic uncertainties, the decay mode $B_s\to\phi\phi$ may be related
through the $SU(3)$ flavour symmetry to $B_s\to \phi \bar K^{*0}$ and plausible 
dynamical assumptions, which can be checked through experimental control 
channels \cite{FG-1}. The current $B$-factory data on the CP asymmetries
of the $b\to s$ penguin modes leave ample space for NP phenomena in
the $B^0_s\to \phi\phi$ decay to be discovered at LHCb. 

In the SM, $B^0_q$--$\bar B^0_q$ mixing ($q\in\{d,s\}$) is governed by box 
diagrams with internal top-quark exchances and is, therefore, a strongly suppressed 
loop phenomenon. In the presence of NP, we may get new contributions through 
NP particles in the box topologies, or new contributions at the tree level. 
In this case, the off-diagonal element of the mass 
matrix is modified as follows \cite{BF-06}:
\begin{equation}\label{M12q}
M_{12}^{(q)} =M_{12}^{q,{\rm SM}} \left(1 + \kappa_q e^{i\sigma_q}\right),
\end{equation}
where the real parameter $\kappa_q$ is a measure of the strength of NP with 
respect to the SM, and $\sigma_q$ denotes a CP-violating NP phase. The mass 
difference $\Delta M_q$ between the two mass eigenstates and the mixing phase 
$\phi_q$ are then modified as
\begin{eqnarray}
\Delta M_q & = &\Delta M_q^{\rm SM}+\Delta M_q^{\rm NP} =
\Delta M_q^{\rm SM}\left| 1 + \kappa_q
  e^{i\sigma_q}\right|,\label{DMq-NP}\\
\phi_q & = & \phi_q^{\rm SM}+\phi_q^{\rm NP}=
\phi_q^{\rm SM} + \arg (1+\kappa_q e^{i\sigma_q}).\label{phiq-NP}
\end{eqnarray}
In the case of the $B^0_d$ mesons, which are accessible at the $B$ factories, 
we have $\phi_d^{\rm SM}=2\beta$. The SM contribution of $\Delta M_d$
depends both on the CKM factor $|V_{td}^\ast V_{tb}|$, which is governed by
$\gamma$ if unitarity is used, and on the hadronic parameter
$f_{B_d}^2\hat B_{B_d}$, which is usually taken from non-perturbative lattice 
QCD calculations \cite{lat}. In particular the measurement of the mixing-induced CP 
violation in $B^0_d\to J/\psi \phi$, which can be converted into
\begin{equation}
\phi_d^{\rm NP}=(2\beta)_{\psi K_{\rm S}}-
(2\beta)_{\rm true}^{\rm tree},
\end{equation}
has a dramatic impact on the allowed region in the $\sigma_d$--$\kappa_d$ plane 
of NP parameters (for a detailed analysis and discussion, see \cite{BF-06}). On the 
other hand, in the case of the $B^0_s$-meson system, we are still left with a large 
allowed region for the corresponding NP parameter space, as we will discuss in the
next section.

\boldmath
\section{$B$ Physics at the LHC}
\unboldmath
The $B$-decay studies at the LHC will allow us to enter a new territory of
the $B$-physics landscape that is still largely unexplored. This is in particular 
due to the high statistics which can quickly be accummulated  and the access 
to the $B_s$-meson system, offering a physics programme that is
to a large extent complementary to that of the $e^+e^-$ $B$ factories 
operated at the $\Upsilon(4S)$ resonance. 

\boldmath
\subsection{General Features of the $B_s$ System}
\unboldmath
In the SM, we expect  a mass difference 
$\Delta M_s={\cal O}(20\,\mbox{ps}^{-1})$, which is much larger than the
experimental value of $\Delta M_d = 0.5\,\mbox{ps}^{-1}$. Consequently,
the $B^0_s$--$\bar B^0_s$ oscillations are very rapid, thereby making it
very challenging to resolve them experimentally. 

Whereas the difference between the decay widths of the mass eigenstates
of the $B^0_d$-meson system is negligible, its counterpart 
$\Delta\Gamma_s/\Gamma_s$ in the $B^0_s$-meson system is expected to 
be of ${\cal O}(10\%)$ \cite{lenz}. Recently, the first results for 
$\Delta\Gamma_s$ were reported from the Tevatron, using the 
$B^0_s\to J/\psi\phi$ channel \cite{DDF}:
\begin{equation}\label{DG-det}
\Delta\Gamma_s=\left\{
\begin{array}{ll}
(0.17\pm0.09\pm0.02)\mbox{ps}^{-1}  & \mbox{(D0 \cite{D0-DG})}\\
(0.076^{+0.059}_{-0.063}\pm0.006)\mbox{ps}^{-1} & \mbox{(CDF \cite{CDF-DG})}.
\end{array}
\right.
\end{equation}
It will be interesting to follow the evolution of these data. At LHCb, we expect a 
precision of $\sigma(\Delta\Gamma_s)=0.027\mbox{ps}^{-1}$ already with
$0.5\,\mbox{fb}^{-1}$ data, which is expected to be available by the end of 2009
\cite{nakada}; ATLAS expects a relative accuracy of $13\%$ with 
$30\,\mbox{fb}^{-1}$ of data taken at low luminosity \cite{smsp}. The width difference 
$\Delta\Gamma_s$ offers studies of CP violation through ``untagged" rates 
of the following form:
\begin{equation}
\langle\Gamma(B_s(t)\to f)\rangle
\equiv\Gamma(B^0_s(t)\to f)+\Gamma(\overline{B^0_s}(t)\to f),
\end{equation}
which are interesting in terms of efficiency, acceptance and purity. If 
both $B^0_s$ and $\bar B^0_s$ states may decay into the final state $f$,
the rapidly oscillating $\Delta M_st$ terms cancel. Various ``untagged" 
strategies exploiting this feature were proposed (see \cite{DDF} and 
\cite{dun,FD-CP,FD-NCP,DFN}); we will discuss an example in 
Section~\ref{ssec:Bspsiphi}.

Finally, the CP-violating phase of $B^0_s$--$\bar B^0_s$ mixing is tiny in the SM:
\begin{equation}\label{phis-SM}
\phi_s^{\rm SM}=-2\lambda^2\eta\approx -2^\circ,
\end{equation}
where $\lambda$ and $\eta$ are the usual Wolfenstein parameters \cite{wolf}.
This feature is very interesting for the search of signals of NP \cite{DFN,NiSi,BMPR} 
(see Section~\ref{ssec:Bspsiphi}).

\begin{figure}[t]
 \centering
  \includegraphics[width=5.8truecm]{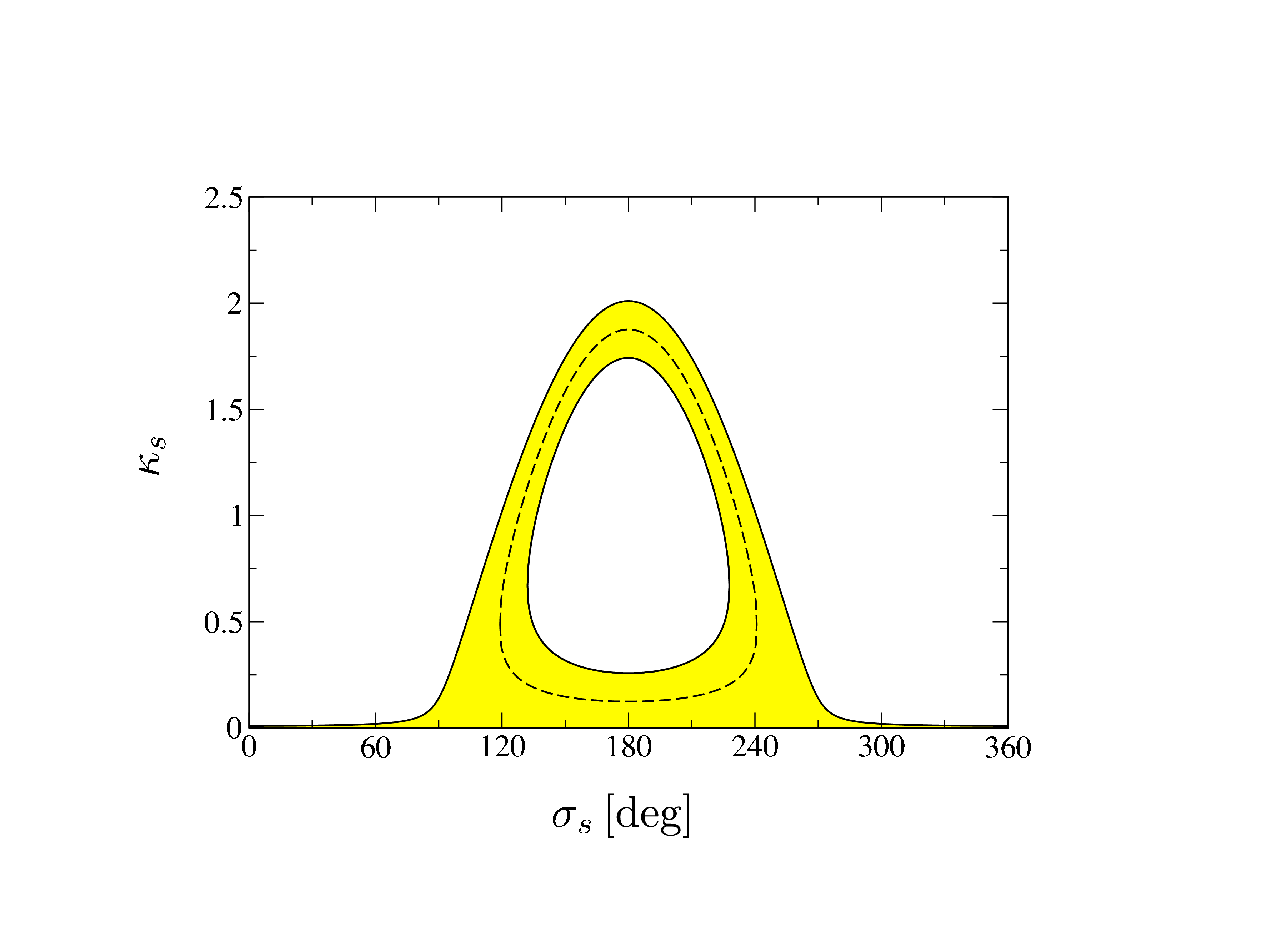} 
     \vspace*{-0.4truecm}
  \caption{The allowed region in the $\sigma_s$--$\kappa_s$ plane of
  NP parameters for $B^0_s$--$\bar B^0_s$ mixing.}\label{fig:MDs-NP}
\end{figure}

\boldmath
\subsection{Measurement of $\Delta M_s$}
\unboldmath
For many years, only lower bounds on $\Delta M_s$ were available from
the LEP (CERN) experiments and SLD (SLAC) \cite{Bosc-WG}. In 2006, the value 
of $\Delta M_s$ could eventually be pinned down at the Tevatron \cite{DMs-obs}.
The most recent results read as follows:
\begin{equation}\label{MDs}
\Delta M_s=\left\{
\begin{array}{ll}
(18.56 \pm0.87){\rm ps}^{-1}  & \mbox{(D0 \cite{D0-DMs})} \\
(17.77\pm0.10 \pm 0.07){\rm ps}^{-1} & 
\mbox{(CDF \cite{CDF-DMs})}.
\end{array}
\right.
\end{equation}
On the other hand, the HPQCD collaboration has reported the following 
lattice QCD prediction \cite{HPQCD-DMs}:
\begin{equation}\label{HPQCD-DMs}
\Delta M_s^{\rm SM}=20.3(3.0)(0.8)\,{\rm ps}^{-1}.
\end{equation}
In contrast to the case of $\Delta M_d$, the CKM factor entering this
SM value does not require information on $\gamma$ and $|V_{ub}/V_{cb}|$,
as 
\begin{equation}
|V_{ts}^*V_{tb}|=|V_{cb}|\left[1+{\cal O}(\lambda^2)\right],
\end{equation}
which is an important advantage. Using (\ref{DMq-NP}) and (\ref{HPQCD-DMs}), 
we may convert the
experimental value of $\Delta M_s$ into the allowed region in the 
$\sigma_s$--$\kappa_s$ plane shown in Fig.~\ref{fig:MDs-NP}, as discussed
in detail in \cite{BF-06}. We see that  the measurement of $\Delta M_s$ leaves 
ample space for the NP parameters $\sigma_s$ and $\kappa_s$, which can
also be accommodated in specific scenarios (e.g.\ SUSY, extra $Z'$ and little 
Higgs models). It should be noted that the experimental errors are already 
significantly smaller than the theoretical lattice QCD uncertainties.
The new experimental results on $\Delta M_s$ have immediately triggered 
a lot of theoretical activity (see, e.g., \cite{BF-06,DMs-papers,BBGT}).

As in the case of the $B_d$-meson system, the allowed region in the 
$\sigma_s$--$\kappa_s$ plane will be dramatically
reduced as soon as measurements of CP violation in the $B_s$-meson 
system become available. The ``golden" channel in this respect is 
$B^0_s\to J/\psi \phi$, our next topic.

\boldmath
\subsection{The Decay $B^0_s\to J/\psi \phi$}\label{ssec:Bspsiphi}
\unboldmath
This mode is the counterpart of the $B^0_d\to J/\psi K_{\rm S}$ transition, where
we have just to replace the down quark by a strange quark. The structures of the
corresponding decay amplitudes are completely analogous to each other. However,
there is also an important difference with respect to $B^0_d\to J/\psi K_{\rm S}$,
since the final state of $B^0_s\to J/\psi \phi$ contains two vector mesons and is,
hence, an admixture of different CP eigenstates. Using the angular distribution of the 
$J/\psi [\to\ell^+\ell^-]\phi [\to\ K^+K^-]$ decay products, the CP eigenstates
can be disentangled \cite{DDLR} and the time-dependent decay rates calculated
\cite{DDF,DFN}. As in the case of $B^0_d\to J/\psi K_{\rm S}$, the
hadronic matrix elements cancel then in the mixing-induced observables. For the
practical implementation, a set of three linear polarization amplitudes is usually 
used: $A_0(t)$ and $A_\parallel(t)$ correspond to CP-even final-state configurations,
whereas $A_\perp(t)$ describes a CP-odd final-state configuration.

It is instructive to illustrate how this works by having a closer look at the
one-angle distribution, which takes the following form \cite{DDF,DFN}:
\begin{equation}
\frac{d\Gamma(B^0_s(t)\to J/\psi \phi)}{d\cos\Theta}\propto
\left(|A_0(t)|^2+|A_\parallel(t)|^2\right)
\frac{3}{8}\left(1+\cos^2\Theta\right)+|A_\perp(t)|^2\frac{3}{4}\sin^2\Theta.
\end{equation}
Here $\Theta$ is defined as the angle between the momentum of the $\ell^+$
and the normal to the decay plane of the $K^+K^-$ system in the $J/\psi$
rest frame. The time-dependent measurement of the angular dependence
allows us to extract the following observables:
\begin{equation}
P_+(t)\equiv |A_0(t)|^2+|A_\parallel(t)|^2, \quad
P_-(t)\equiv |A_\perp(t)|^2,
\end{equation}
where $P_+(t)$ and $P_-(t)$ refer to the CP-even and CP-odd final-state configurations,
respectively. If we consider the case of having an initially, i.e.\ at time $t=0$, present
$\bar B^0_s$ meson, the CP-conjugate quantities $\bar P_\pm(t)$ can be extracted
as well. Using an {\it untagged} data sample, the untagged rates
\begin{equation}
P_\pm(t)+\overline{P}_\pm(t)\propto
\left[(1\pm\cos\phi_s)e^{-\Gamma_{\rm L}t}+
(1\mp\cos\phi_s)e^{-\Gamma_{\rm H}t}\right]
\end{equation}
can be determined, while a {\it tagged} data sample allows us to measure
the CP-violating asymmetries
\begin{equation}
\frac{P_\pm(t)-\overline{P}_\pm(t)}{P_\pm(t)+\overline{P}_\pm(t)}=
\pm\left[\frac{2\,\sin(\Delta M_st)\sin\phi_s}{(1\pm\cos\phi_s)e^{+\Delta\Gamma_st/2}+
(1\mp\cos\phi_s)e^{-\Delta\Gamma_st/2}}\right].
\end{equation}
In the presence of CP-violating NP contributions to $B^0_s$--$\bar B^0_s$
mixing, we obtain
\begin{equation}
\phi_s=-2\lambda^2\eta+\phi_s^{\rm NP}\approx -2^\circ+
\phi_s^{\rm NP}\approx \phi_s^{\rm NP}.
\end{equation}
Consequently, NP of this kind would be indicated by the following features:
\begin{itemize}
\item The {\it untagged} observables depend on {\it two} exponentials;
\item {\it sizeable} values of the CP-violating asymmetries.
\end{itemize}

\begin{figure}[t] 
 \centering
\begin{tabular}{cc}
  \includegraphics[width=5.8truecm]{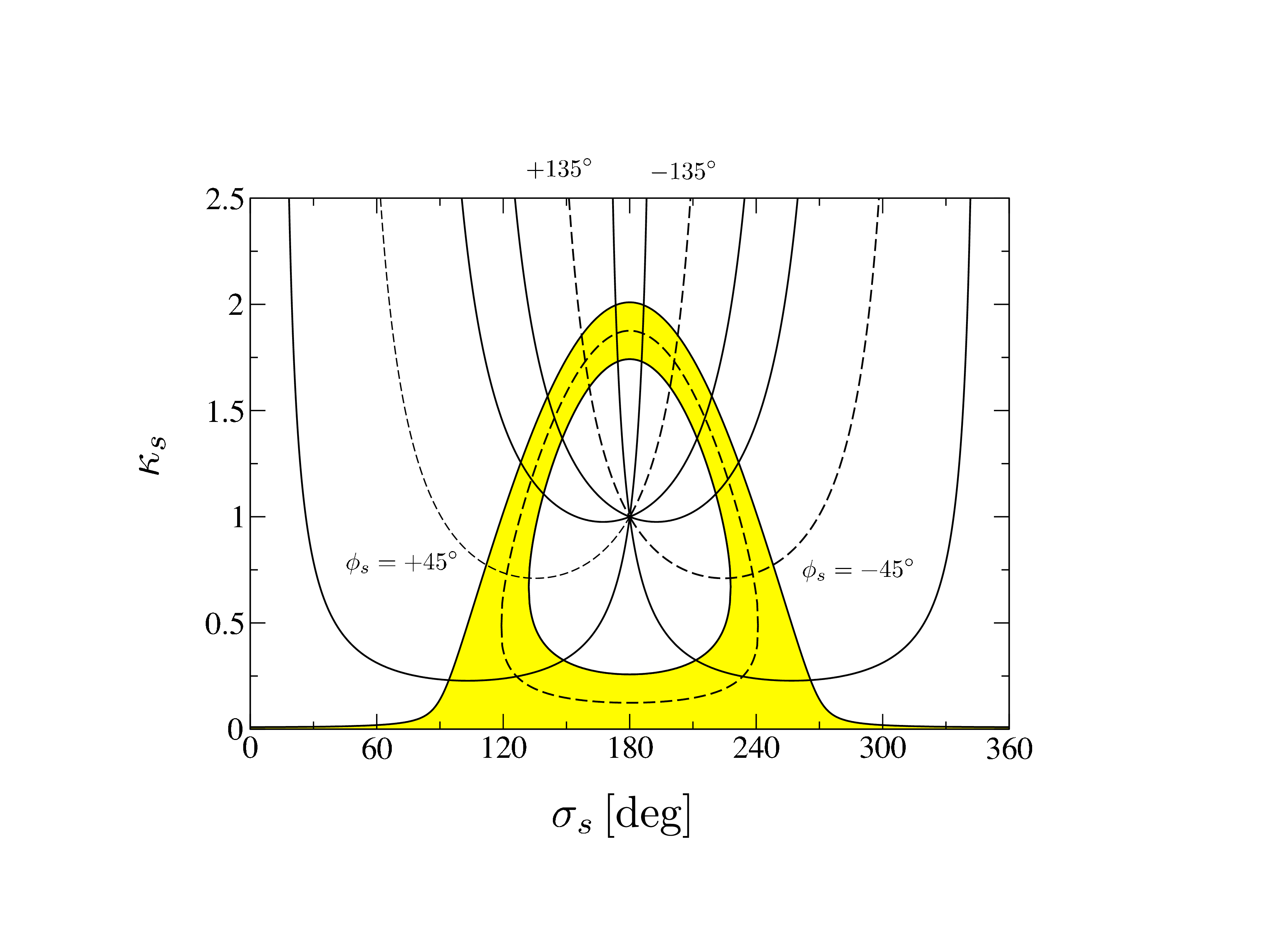} &
    \includegraphics[width=5.8truecm]{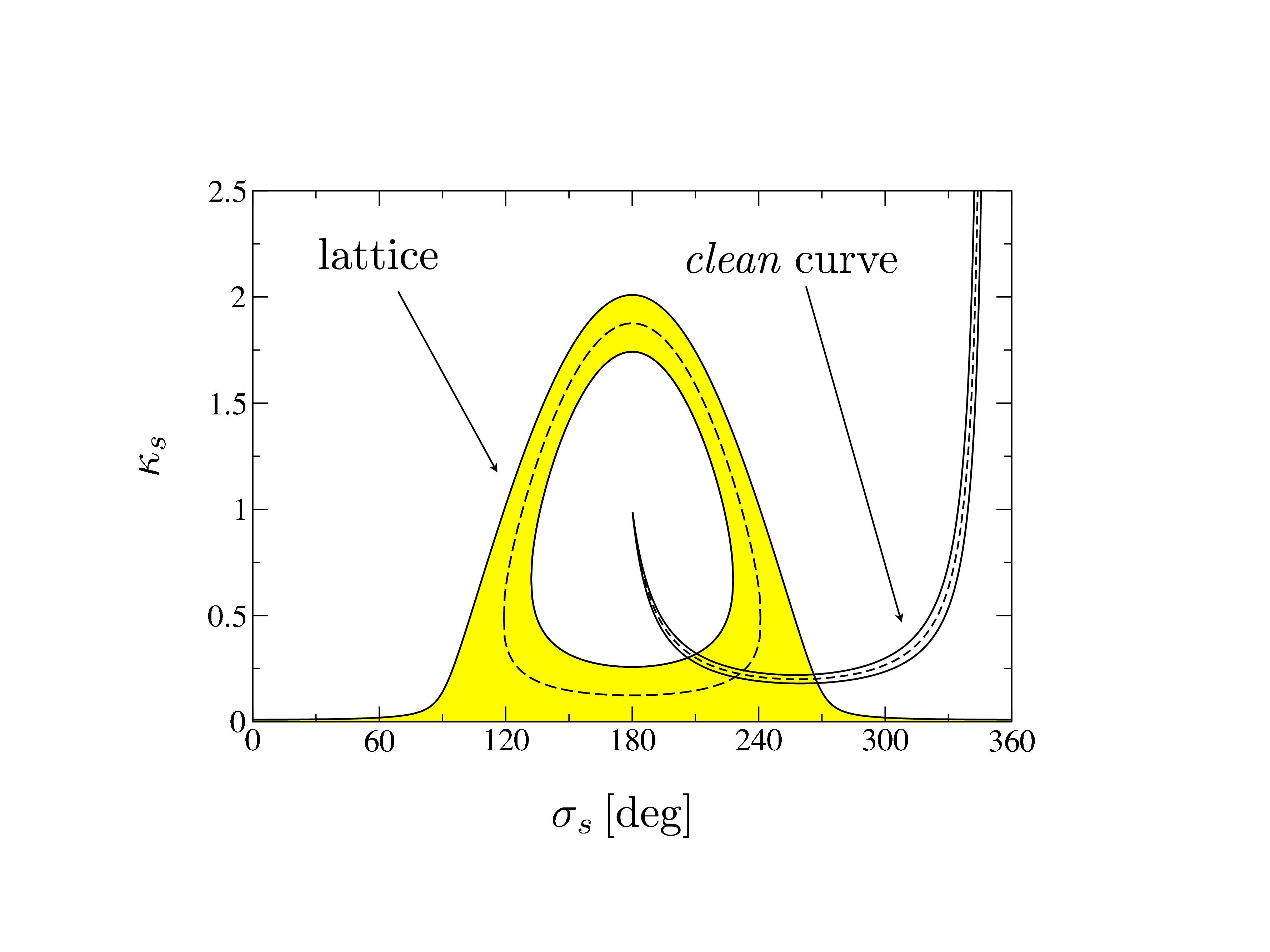} 
    \end{tabular}
    \vspace*{-0.4truecm}
   \caption[]{Impact of the measurement of CP violation in 
   $B^0_s\to J/\psi\phi$: current D0 data [left panel], and a NP scenario
   with $(\sin\phi_s)_{\rm exp}=-0.20\pm0.02$ [right panel].}\label{fig:sis-kas-CP}
\end{figure}

These general features hold also for the full three-angle distribution
\cite{DDF,DFN}: it is much more involved than the one-angle case, but
provides also additional information through interference terms of the
form 
\begin{equation}
\mbox{Re}\{A_0^\ast(t)A_\parallel(t)\}, \quad
\mbox{Im}\{A_f^\ast(t)A_\perp(t)\} \, (f\in\{0,\parallel\}).
\end{equation}
From an experimental point of view, there is no experimental draw-back with
respect to the one-angle case. Following these lines, $\Delta\Gamma_s$ 
(see (\ref{DG-det})) and $\phi_s$ can be extracted.  
Recently, the D0 collaboration has reported first results for the measurement 
of $\phi_s$ through the untagged, time-dependent three-angle 
$B^0_s\to J/\psi\phi$ distribution
\cite{D0-phis}:
\begin{equation}
\phi_s=-0.79\pm0.56\,\mbox{(stat.)} ^{+0.14}_{-0.01}\,\mbox{(syst.)}
=-(45\pm32^{+1}_{-8})^\circ,
\end{equation}
which is complemented by three additional mirror solutions. This phase is therefore not 
yet stringently constrained. Using (\ref{phiq-NP}), we then obtain the curves 
in the $\sigma_s$--$\kappa_s$ plane shown in the left panel of
Fig.~\ref{fig:sis-kas-CP}. Very recently, the CDF collaboration reported first
bounds on $\phi_s$ from flavour-tagged $B^0_s\to J/\psi\phi$
decays \cite{cdf-tagged}.

Fortunately, $\phi_s$ will be very accessible at LHCb, where already the initial
$0.5\,\mbox{fb}^{-1}$ of data will give an uncertainty of
$\sigma(\phi_s)=0.046=2.6^\circ$ by the end of 2009, which will
be significantly improved further thanks to the $2\,\mbox{fb}^{-1}$ that should
be available by the end of 2010 \cite{nakada}. At some point, also in view of LHCb upgrade plans \cite{LHCb-up}, we have to include hadronic penguin uncertainties. 
This can be done with the help of the $B^0_d\to J/\psi \rho^0$ decay \cite{RF-ang}. In
order to illustrate the impact of the measurement of CP violation in 
$B^0_s\to J/\psi\phi$, we show in the right panel of Fig.~\ref{fig:sis-kas-CP}
the case corresponding to $(\sin\phi_s)_{\rm exp}=-0.20\pm0.02$. Such a 
measurement would give a NP signal at the $10\,\sigma$ level and demonstrates 
the power of the $B_s$ system to search for NP \cite{BF-06}.  It should be
emphasized that the contour following from the measurement of
$\phi_s$ would be essentially clean, in contrast to the shaded region representing 
the constraint from the measured value of $\Delta M_s$, which suffers from 
lattice QCD uncertainties.

\subsection{Further Benchmark Decays for LHCb}
This experiment has a very rich physics programme. Besides many other 
interesting aspects, there are two major lines of research:
\begin{enumerate}
\item Precision measurements of $\gamma$:\\
On the one hand, there are strategies using  tree decays: 
$B^0_s\to D_s^\mp K^\pm$ [$\sigma_\gamma\sim5^\circ$],
$B^0_d\to D^0K^{*}$ [$\sigma_\gamma\sim8^\circ$],
$B^\pm\to D^0K^\pm$ [$\sigma_\gamma\sim5^\circ$],
where we have also indicated the expected sensitivities for $10\,\mbox{fb}^{-1}$;
by 2013, a LHCb tree deterimation of $\gamma$ with 
$\sigma_\gamma=2^\circ\sim3^\circ$
should be available \cite{nakada}. This very impressive situation should be
compared with the current $B$-factory data, yielding
\begin{equation}
\left.\gamma\right|_{D^{(*)} K^{(*)}} = \left\{
\begin{array}{ll}
(77^{+30}_{-32})^\circ & \mbox{(CKMfitter \cite{CKMfitter}),}\\[5pt] 
(88\pm 16)^\circ & \mbox{(UTfit \cite{UTfit}).}
\end{array}
\right.
\end{equation}
These extractions are essentially unaffected by NP effects.
On the other hand, $\gamma$ can also be determined through $B$ decays
with penguin contributions: $B^0_s\to K^+K^-$ and $B^0_d\to \pi^+\pi^-$
[$\sigma_\gamma\sim5^\circ$], $B^0_s\to D_s^+D_s^-$ and $B^0_d\to D_d^+D_d^-$.
The key question is whether discrepancies will arise in these determinations. 

\item Analyses of rare decays, which are absent at the SM tree level:\\ 
prominent examples are $B^0_{s,d}\to\mu^+\mu^-$,
$B^0_d\to K^{*0}\mu^+\mu^-$ and $B^0_s\to \phi \mu^+\mu^-$. In order to 
complement the studies of CP violation in $b\to s$ penguin modes at the
$B$ factories, $B^0_s\to\phi\phi$ is a very interesting mode for LHCb, as noted in
Section~\ref{sec:2}.   
\end{enumerate}
Let us next have a closer look at some of these decays.

\boldmath
\subsubsection{CP Violation in $B_s\to D_s^\pm K^\mp$ and $B_d\to D^\pm\pi^\mp$}
\unboldmath
The pure tree decays $B_s\to D_s^\pm K^\mp$ \cite{BsDsK} and 
$B_d\to D^\pm \pi^\mp$ \cite{BdDpi} can be treated on the same theoretical 
basis, and provide new strategies to determine $\gamma$ \cite{RF-gam-ca}. 
Following this paper, we write these modes as $B_q\to D_q \bar u_q$. Their
characteristic feature is that both a $B^0_q$ and a $\bar B^0_q$ meson may decay 
into the same final state $D_q \bar u_q$. Consequently,  interference effects 
between $B^0_q$--$\bar B^0_q$ mixing and decay processes arise, which 
involve the CP-violating phase combination $\phi_q+\gamma$.

In the case of $q=s$, i.e.\ $D_s\in\{D_s^+, D_s^{\ast+}, ...\}$ and 
$u_s\in\{K^+, K^{\ast+}, ...\}$, these interference effects are governed 
by a hadronic parameter $X_s e^{i\delta_s}\propto R_b\approx0.4$, where
$R_b\propto |V_{ub}/V_{cb}|$ is the usual UT side, and hence are large. 
On the other hand, for $q=d$, i.e.\ $D_d\in\{D^+, D^{\ast+}, ...\}$ 
and $u_d\in\{\pi^+, \rho^+, ...\}$, the interference effects are described 
by $X_d e^{i\delta_d}\propto -\lambda^2R_b\approx-0.02$, and hence are tiny. 

Measuring the $\cos(\Delta M_qt)$ and $\sin(\Delta M_qt)$ terms of the
time-dependent $B_q\to D_q \bar u_q$ rates, a theoretically clean 
determination of $\phi_q+\gamma$ is possible \cite{BsDsK,BdDpi}. 
Since the $\phi_q$ can be determined separately, $\gamma$ 
can be extracted. However, in the practical implementation, there are problems:
we encounter an eightfold discrete ambiguity for $\phi_q+\gamma$, which
is very disturbing for the search of NP, and in the $q=d$ case, an additional input 
is required to extract $X_d$ since
${\cal O}(X_d^2)$ interference effects would otherwise have to be resolved,
which is impossible. Performing a combined analysis of the $B^0_s\to D_s^{+}K^-$
and $B^0_d\to D^+\pi^-$ decays, these problems can be solved \cite{RF-gam-ca}.
This strategy exploits the fact that these transitions are related to each other
through the $U$-spin symmetry of strong interactions,\footnote{The $U$ spin is an
$SU(2)$ subgroup of the $SU(3)_{\rm F}$ flavour-symmetry group of QCD, connecting
$d$ and $s$ quarks in analogy to the isospin symmetry, which relates $d$ and
$u$ quarks to each other.} allowing us to simplify the hadronic sector. Following
these lines, an unambiguous value of $\gamma$ can be extracted from the
observables. To this end, $X_d$ has actually not to be fixed, and $X_s$ may only enter
through a $1+X_s^2$ correction, which is determined through untagged $B_s$
rates. The first studies for LHCb are very promising, and are 
currently further refined \cite{WG5-rep}.

\boldmath
\subsubsection{The $B_s\to K^+K^-$, $B_d\to \pi^+\pi^-$ System}
\begin{figure}[t]
   \centering
  \includegraphics[width=7.0truecm]{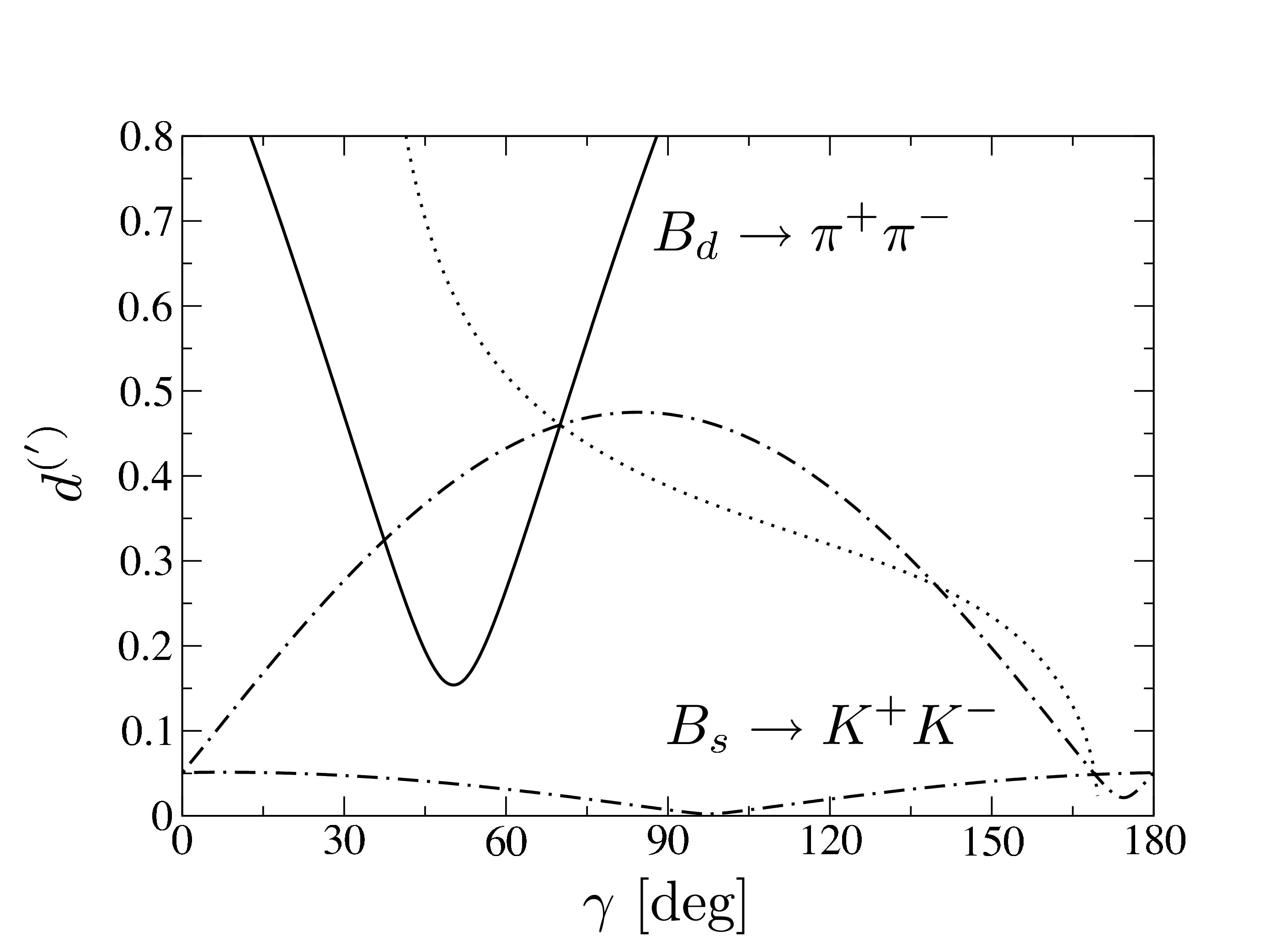} 
  \vspace*{-0.3truecm}
     \caption{The contours in the $\gamma$--$d^{(')}$ plane for an example 
     corresponding to the CP asymmetries
   ${\cal A}_{\rm CP}^{\rm dir}(B_d\to\pi^+\pi^-)=-0.24$ and 
   ${\cal A}_{\rm CP}^{\rm mix}(B_d\to\pi^+\pi^-)=+0.59$, as well as
   ${\cal A}_{\rm CP}^{\rm dir}(B_s\to K^+K^-)=+0.09$ and
   ${\cal A}_{\rm CP}^{\rm mix}(B_s\to K^+K^-)=-0.23$.}\label{fig:Bs-Bd-contours}
\end{figure}

\unboldmath
The decay $B^0_s\to K^+K^-$ is a $\bar b \to \bar s$ transition, and
involves tree and penguin amplitudes, as $B^0_d\to\pi^+\pi^-$ \cite{RF-BsKK,FlMa}. 
However, because of the different CKM structure, the latter 
topologies play actually the dominant r\^ole in $B^0_s\to K^+K^-$,
whereas the major contribution to $B^0_d\to\pi^+\pi^-$ is due to the tree
amplitude.  In the SM, we may write
\begin{eqnarray}
A(B^0_d\to\pi^+\pi^-)& \propto & \left[e^{i\gamma}-de^{i\theta}\right]\\
A(B_s^0\to K^+K^-)& \propto &
\left[e^{i\gamma}+\left(\frac{1-\lambda^2}{\lambda^2}\right)d'e^{i\theta'}\right],
\end{eqnarray}
where the CP-conserving hadronic parameters $de^{i\theta}$ and
$d'e^{i\theta'}$ descripe -- sloppily speaking -- the ratios of penguin to tree
contributions. The direct and mixing-induced CP asymmetries take then the 
following general form:
\begin{equation}
{\cal A}_{\rm CP}^{\rm dir}(B_d\to \pi^+\pi^-)=
G_1(d,\theta;\gamma), \quad
{\cal A}_{\rm CP}^{\rm mix}(B_d\to \pi^+\pi^-)=
G_2(d,\theta;\gamma,\phi_d)
\end{equation}
\begin{equation}
{\cal A}_{\rm CP}^{\rm dir}(B_s\to K^+K^-)=
G_1'(d',\theta';\gamma), \quad
{\cal A}_{\rm CP}^{\rm mix}(B_s\to K^+K^-)=
G_2'(d',\theta';\gamma,\phi_s).
\end{equation}
Since $\phi_d$ is already known  and  $\phi_s$ is negligibly small
in the SM -- or can be determined through $B^0_s\to J/\psi \phi$ should CP-violating
NP contributions to $B^0_s$--$\bar B^0_s$ mixing make it sizeable -- 
we may convert the measured values of 
${\cal A}_{\rm CP}^{\rm dir}(B_d\to \pi^+\pi^-)$, 
${\cal A}_{\rm CP}^{\rm mix}(B_d\to \pi^+\pi^-)$ and
${\cal A}_{\rm CP}^{\rm dir}(B_s\to K^+K^-)$, 
${\cal A}_{\rm CP}^{\rm mix}(B_s\to K^+K^-)$ into {\it theoretically clean}
contours in the $\gamma$--$d$ and $\gamma$--$d'$ planes, respectively.
In Fig.~\ref{fig:Bs-Bd-contours}, we show these contours (solid and dot-dashed) 
for an example, which is inspired by the current $B$-factory data.

A closer look at the corresponding Feynman diagrams shows that 
$B^0_d\to\pi^+\pi^-$ is actually related to $B^0_s\to K^+K^-$ through the interchange 
of all down and strange quarks. Consequently, each decay topology contributing
to $B^0_d\to\pi^+\pi^-$ has a counterpart in $B^0_s\to K^+K^-$ and vice versa, 
and the corresponding hadronic parameters can be related to each other
with the help of the $U$-spin flavour symmetry of strong interactions,
implying the following relations \cite{RF-BsKK}:
\begin{equation}\label{U-spin-rel}
d'=d, \quad \theta'=\theta.
\end{equation}
Applying the former, we may extract $\gamma$ and $d$ through the 
intersections of the theoretically clean $\gamma$--$d$ and $\gamma$--$d'$ 
contours. In the example of Fig.~\ref{fig:Bs-Bd-contours}, a twofold
ambiguity arises from the solid and dot-dashed curves. However, as 
discussed in \cite{RF-BsKK}, it can be resolved with the help of the dotted 
contour, thereby leaving us with the ``true" solution of $\gamma=70^\circ$ in this 
case. Moreover, we may determine $\theta$ and $\theta'$, which allow an interesting 
internal consistency check of the second $U$-spin relation in (\ref{U-spin-rel}).

This strategy is very promising from an experimental point of view for LHCb, 
where an accuracy for $\gamma$ of a few degrees can be achieved 
\cite{LHCb-analyses}. As far as possible $U$-spin-breaking 
corrections to $d'=d$ are concerned, they enter the determination of $\gamma$ 
through a relative shift of the $\gamma$--$d$ and $\gamma$--$d'$ contours; 
their impact on the extracted value of $\gamma$ therefore depends on the form 
of these curves, which is fixed through the measured observables. In the examples discussed in \cite{RF-BsKK} and Fig.~\ref{fig:Bs-Bd-contours}, the extracted value 
of $\gamma$ would be very stable with respect to such effects. It should also be noted
that the $U$-spin relations in (\ref{U-spin-rel}) are particularly robust since they 
involve only ratios of hadronic amplitudes, where all $SU(3)$-breaking decay constants
and form factors cancel in factorization and also chirally enhanced terms
would not lead to  $U$-spin-breaking corrections \cite{RF-BsKK}.

\begin{figure}
\centerline{
 \includegraphics[width=7.0truecm]{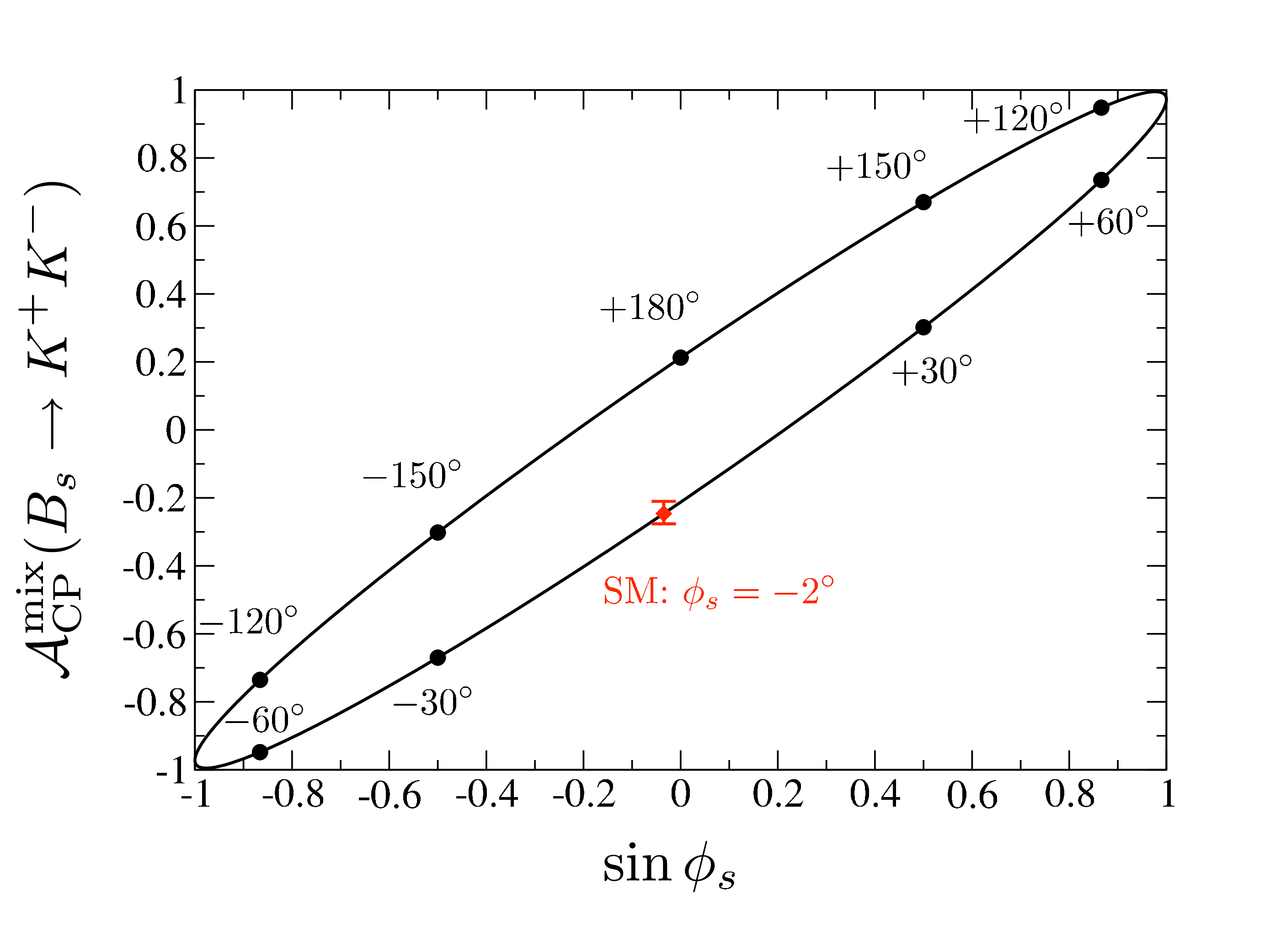}}
 \vspace*{-0.3truecm}
\caption{The correlation between $\sin\phi_s$, which can be determined 
through mixing-induced CP violation in $B^0_s\to J/\psi \phi$, and 
${\cal A}_{\rm CP}^{\rm mix}(B_s\to K^+K^-)$. Each point on the curve 
corresponds to a given value of $\phi_s$, as indicated by the numerical 
values \cite{RF-BsKK-07}.}\label{fig:Bs-NP-2}
\end{figure}

A detailed analysis of the status and prospects of the $B_{s,d}\to \pi\pi,\pi K, KK$ 
system in view of the first results on the $B_s$ modes from the Tevatron 
\cite{CDF-BsK+K-} was performed in \cite{RF-BsKK-07}. Interestingly, the data 
for the  $B_d\to\pi^+\pi^-$, $B_s\to K^+K^-$ system favour the BaBar measurement 
of the direct CP violation in $B_d\to\pi^+\pi^-$, which results in a fortunate situation
for the extraction of $\gamma$, yielding $\gamma=(66.6^{+4.3+4.0}_{-5.0-3.0})^\circ$, 
where the latter errors correspond to a an estimate of $U$-spin-breaking effects. An important further step will be the measurement of mixing-induced CP violation in 
$B_s\to K^+K^-$, which is predicted in the SM as 
${\cal A}_{\rm CP}^{\rm mix}(B_s\to K^+K^-)=-0.246^{+0.036+0.052}_{-0.030-0.024}$, 
where the second errors illustrate the impact of large non-factorizable 
$U$-spin-breaking corrections. In the case of CP-violating NP contributions to 
$B^0_s$--$\bar B^0_s$ mixing also this observable
would be sensitively affected, as can be seen in Fig.~\ref{fig:Bs-NP-2}, and
allows an unambiguous determination of  the $B^0_s$--$\bar B^0_s$ mixing 
phase with the help of $B_s\to J/\psi \phi$ at LHCb. Finally, the measurement 
of direct CP violation in $B_s\to K^+K^-$ will make the full exploitation 
of the physics potential of the $B_{s,d}\to \pi\pi, \pi K, KK$ modes possible.

\boldmath
\subsubsection{The Rare Decays $B_{s,d}\to\mu^+\mu^-$}
\unboldmath
In the SM, these decays originate from $Z$ penguins and box diagrams.
A closer look at the corresponding low-energy effective Hamiltonian \cite{BBL}
shows that the hadronic matrix element is simply given by the decay constant 
$f_{B_q}$. Consequently, we arrive at a very favourable situation with respect 
to the hadronic matrix elements. Since, moreover, NLO QCD corrections were 
calculated, and long-distance contributions are expected to play a negligible 
r\^ole \cite{Bmumu}, the $B^0_q\to\mu^+\mu^-$ modes belong to the cleanest 
rare $B$ decays. 

Using also the data for the mass differences $\Delta M_q$ to reduce the hadronic 
uncertainties,\footnote{This input allows us to replace the decay constants $f_{B_q}$
through the bag parameters $\hat B_{B_q}$.} the following SM predictions were 
obtained in \cite{BBGT}:
\begin{eqnarray}
\mbox{BR}(B_s\to\mu^+\mu^-) &=& (3.35\pm0.32)\times 10^{-9}\\
\mbox{BR}(B_d\to\mu^+\mu^-) &=& (1.03\pm0.09)\times 10^{-10}.
\end{eqnarray}
Consequently, these branching ratios are extremely tiny. But things could actually
be much more exciting, as NP effects may significantly enhance 
$\mbox{BR}(B_s\to\mu^+\mu^-)$. The current upper bounds (95\% C.L.) from 
the CDF collaboration read as follows
\cite{CDF-Bmumu}:
\begin{equation}\label{Bmumu-exp-CDF}
\mbox{BR}(B_s\to\mu^+\mu^-)<5.8\times10^{-8}, \quad
\mbox{BR}(B_d\to\mu^+\mu^-)<1.8 \times10^{-8},
\end{equation}
while the D0 collaboration finds the following 90\% C.L. (95\% C.L.) upper limit 
\cite{D0-Bmumu}:
\begin{equation}\label{Bmumu-exp-D0}
\mbox{BR}(B_s\to\mu^+\mu^-)<7.5~(9.3) \times 10^{-8}.
\end{equation}
Consequently, there is still a long way to go within the SM. However, by the end
of 2009, with $0.5\,\mbox{fb}^{-1}$ data, LHCb can exclude or discover a 
NP contribution to $B_s\to\mu^+\mu^-$  at the level of the SM \cite{nakada}. 
This decay is also very interesting for ATLAS and CMS, where detailed signal
and background studies are currently in progress \cite{smsp}.

\boldmath
\subsubsection{The Rare Decay $B^0_d\to K^{*0}\mu^+\mu^-$}
\unboldmath
The key observable for NP searches through this channel is the following
forward--backward asymmetry:
\begin{equation}
A_{\rm FB}(\hat s)=\frac{1}{{\rm d}\Gamma/{\rm d}\hat s}
\left[\int_0^{+1} {\rm d}(\cos\theta)\frac{{\rm d}^2\Gamma}{{\rm d} \hat s \, 
{\rm d}(\cos\theta)} - \int_{-1}^0{\rm d}(\cos\theta)
\frac{{\rm d}^2\Gamma}{{\rm d} \hat s \, {\rm d}(\cos\theta)}\right].
\end{equation}
Here $\theta$ is the angle between the $B^0_d$ momentum and that of the 
$\mu^+$ in the dilepton centre-of-mass system, and $\hat s \equiv s/M_B^2$ with 
$s=(p_{\mu^+}+p_{\mu^-})^2$. A particularly interesting kinematical point is
characterized by  
\begin{equation}
A_{\rm FB}(\hat s_0)|_{\rm SM}=0,
\end{equation}
as $\hat s_0$ is quite robust with respect to hadronic uncertainties 
(see, e.g., \cite{BKastll}). In SUSY extensions of the SM, $A_{\rm FB}(\hat s)$ 
could take opposite sign or take a dependence on $\hat s$ without a zero point 
\cite{ABHH}. The current $B$-factory data for the inclusive $b\to s\ell^+\ell^-$ 
branching ratios and the integrated forward--backward asymmetries are in 
accordance with the SM, but suffer still from large uncertainties. This situation will 
improve dramatically at the LHC. Here LHCb will measure the zero crossing
point by $\sim2013$ with $10\,\mbox{fb}^{-1}$ with 
$\sigma(s_0)=0.27 (\mbox{GeV}/c^2)^2$,
corresponding to 19k events \cite{nakada}. For other interesting observables 
provided by  $B^0_d\to K^{*0}\mu^+\mu^-$, see \cite{matias-rare}. Also 
alternative $b\to s\mu^+\mu^-$ modes are currently under study, such as 
$B^0_s\to\phi\mu^+\mu^-$ and $\Lambda_b\to \Lambda\mu^+\mu^-$
\cite{nakada,smsp}.

\section{Conclusions and Outlook}
We have seen tremendous progress in $B$ physics during the recent years,
which was made possible through a fruitful interplay between theory and
experiment. Altogether, the $e^+e^-$ $B$ factories have already produced 
${\cal O}(10^9)$ $B\bar B$ pairs, and the Tevatron has recently succeeded in 
observing $B^0_s$--$\bar B^0_s$ mixing. The data agree globally with the KM 
mechanism of CP violation in an impressive manner, but we have also hints for 
discrepancies, which could be first signals of NP. Unfortunately, definite conclusions 
cannot yet be drawn as the uncertainties are still too large. 

Exciting new perspectives for $B$ physics and the exploration of CP violation 
will arise in the summer of 2008 through the start of the LHC, with its
dedicated $B$-decay experiment LHCb. Thanks to the large statistics that
can be collected there and the full exploitation of the physics potential of the
$B_s$-meson system, we will be able to enter a new territory in the exploration
of CP violation. The golden channel for the search of CP-violating NP 
contributions to $B^0_s$--$\bar B^0_s$ mixing is $B^0_s\to J/\psi \phi$, where the
recent measurement of $\Delta M_s$ still leaves ample space for such effects both in 
terms of the general NP parameters and in specific extensions of the SM. In contrast
to the theoretical interpretation of $\Delta M_s$, the corresponding CP asymmetries
have not to rely on non-perturbative lattice QCD calculations. Moreover, it will
be very interesting to search for CP-violating NP effects in $b\to s$ penguin
processes through the $B^0_s\to \phi\phi$ channel. These measurements will
be complemented by other key ingredients for the search of NP: precision 
measurements of the UT angle $\gamma$ by means of various processes with 
a very different dynamics, and powerful analyses of rare $B$ decays. 

In addition to $B$ physics, there are other important flavour probes. An outstanding
example is charm physics, where evidence for $D^0$--$\bar D^0$ mixing was
found at the $B$ factories in the spring of 2007 \cite{D-mix}, and
very recently also at CDF \cite{CDF-Dmix}. The mixing parameters
are measured in the ball park of the SM predictions, which are unfortunately affected
by large long-distance effects. A striking NP signal would be given by CP-violating
effects (for recent theoretical analyses, see, e.g.\ \cite{D-th}). There is also a
powerful charm-physics programme at LHCb. As far as kaon physics is concerned,
the future is given by the rare decays $K^+\to\pi^+\nu\bar\nu$ and 
$K_{\rm L}\to \pi^0\nu\bar\nu$, which are theoretically very clean and would be
very desirable to be measured (efforts at CERN and KEK/J-PARC). Moreover, 
there is of course also exciting flavour violation in the lepton sector (neutrino 
physics, search for $\mu\to e\gamma$ at MEG, etc.), and it is crucial to obtain 
eventually the whole picture. 

The main goal of the ATLAS and CMS experiments is to explore electroweak 
symmetry breaking, in particular the question of whether this is actually caused by 
the Higgs mechanism, to produce and observe new particles, and then to go
back to the deep questions of particle physics, such as the origin of dark matter
and the baryon asymmetry of the Universe. It is obvious that there will be a 
fruitful interplay between these ``direct" studies of NP and the ``indirect" 
information provided by flavour physics, including the $B$-meson system, 
but also $D$, $K$,  and top physics as well as the flavour physics in the lepton 
sector \cite{CERN-WS}. I have no doubts that the next years will be extremely exciting!

\bigskip
I would like to thank Stephan Narison for the invitation to this most interesting 
conference in Madagascar!

\end{document}